\documentclass[showpacs, oneside, twocolumn, prd, amsmath, amssymb, nofootinbib, superscriptaddress]{revtex4-1}

\usepackage{cases}
\usepackage{amsmath}
\usepackage{amssymb}
\usepackage{amsfonts}
\usepackage{amssymb}
\usepackage{dcolumn}
\usepackage{bm}
\usepackage{bbm}
\usepackage{graphicx}
\usepackage{xcolor}
\usepackage{array}
\usepackage{subfigure}
\usepackage{hyperref}
\usepackage{wasysym}
\usepackage{mathtools}
\usepackage{cleveref}
\usepackage{tikz}
\usepackage{relsize}
\usepackage[compat=1.0.0]{tikz-feynman}
\hypersetup{colorlinks=true, urlcolor=black, linkcolor=blue, citecolor=red} %
\allowdisplaybreaks[4]

\begin{document}

\title{Missing one-loop contributions in secondary gravitational waves}

\author{Chao Chen}
\email{iascchao@ust.hk}
\affiliation{Jockey Club Institute for Advanced Study, The Hong Kong University of Science and Technology, Clear Water Bay, Kowloon, Hong Kong, People's Republic of China}

\author{Atsuhisa Ota}
\email{iasota@ust.hk}
\affiliation{Jockey Club Institute for Advanced Study, The Hong Kong University of Science and Technology, Clear Water Bay, Kowloon, Hong Kong, People's Republic of China}

\author{Hui-Yu Zhu}
\email{hzhuav@connect.ust.hk}

\author{Yuhang Zhu}
\email{yzhucc@connect.ust.hk}
\affiliation{Jockey Club Institute for Advanced Study, The Hong Kong University of Science and Technology, Clear Water Bay, Kowloon, Hong Kong, People's Republic of China}
\affiliation{Department of Physics, The Hong Kong University of Science and Technology, Clear Water Bay, Kowloon, Hong Kong, People's Republic of China}

\begin{abstract}

We find several missing one-loop-order contributions in previous considerations about secondary gravitational waves induced at nonlinear order in cosmological perturbations. We consider a consistent perturbative expansion to third-order in cosmological perturbations, including higher-order interactions and iterative solutions ignored in the previous literature.
Tensor fluctuations induced by the source with two scalar and one tensor perturbations are correlated with the first-order tensor fluctuation and thus give a one-loop-order correction to the tensor power spectrum.
The missing loop correction is \textit{scale-invariant} and \textit{negative} in the superhorion region, which secondarily reduces the initial primordial tensor power spectrum prior to the horizon re-entry.
Such an IR behavior is very different from the auto-spectrum of second-order induced tensor modes discussed in the previous literature and can be important for the actual gravitational wave measurements.
For a sharp peak of scalar fluctuations with $A_\zeta=10^{-2}$ at $k_*=10^{5}h/{\rm Mpc}$ motivated by the LIGO/Virgo events, we show that the tensor power spectrum at the cosmic microwave background scale reduces by at most 35\%.
Hence, the polarization B-mode might not be seen because of the reduction of the original tensor spectrum due to the secondary effect of primordial black hole formation.

\end{abstract}

\maketitle

\section{Introduction}

The direct detections of gravitational waves (GWs) from the binary black hole/neutron star mergers break new ground in physical cosmology \cite{LIGOScientific:2016aoc, LIGOScientific:2017vwq}, marking a new era of the multi-messenger astronomy by combining GWs, electromagnetic and neutrino observations \cite{LIGOScientific:2017ync}. The interaction between GWs and matter is weak.
Hence, GWs propagate almost freely through the Universe and carry unique astrophysical and cosmological information. 
The primordial gravitational waves (PGWs) produced in the very early Universe are generally predicted by various early cosmology scenarios \cite{Guth:1980zm,Linde:1981mu,Starobinsky:1980te, Albrecht:1982wi,  Cai:2011tc, Brandenberger:2016vhg}. Currently, the B-mode polarization of cosmic microwave background (CMB) radiation is a promising channel to detect PGWs \cite{CMB-S4:2020lpa}, which may enable us to test the origin of the Universe soon.

In recent years, the secondary GWs induced by the nonlinear coupling of scalar perturbations have been attracting great attention.
Those are regarded as a reasonable tool to detect a type of ultra-compact objects that may exist in the early Universe ---primordial black holes (PBHs) \cite{Ananda:2006af, Baumann:2007zm, Saito:2008jc, Inomata:2016rbd, Kohri:2018awv, Bartolo:2018rku, Cai:2019jah, Garcia-Bellido:2017aan, Inomata:2020cck, Domenech:2021ztg, Cai:2021wzd, Ota:2022hvh}, and to probe the statistical properties of the small-scale primordial curvature perturbations \cite{Cai:2018dig, Unal:2018yaa, Inomata:2018epa, Ota:2020vfn, Atal:2021jyo, Dimastrogiovanni:2022eir, Chen:2022qec}. Overdense regions in the early Universe may stop expanding and collapse to form PBHs \cite{ Hawking:1971ei,Carr:1974nx,Carr:1975qj}.
Sufficiently large density fluctuations for PBH formation can be realized in various inflationary models, e.g., the ultra-slow-roll phase \cite{Garcia-Bellido:2017mdw, Germani:2017bcs, Byrnes:2018txb, Liu:2020oqe, Fu:2020lob, Liu:2020oqe, Inomata:2021tpx, Tasinato:2020vdk, Ozsoy:2021pws, Cole:2022xqc}, the extra fields \cite{Kohri:2012yw, Kawasaki:2012wr, Pi:2017gih, Anguelova:2020nzl, Palma:2020ejf,Fumagalli:2020adf, Cai:2021wzd, Pi:2021dft}, the non-Gaussianity \cite{Ezquiaga:2019ftu, Atal:2019erb, Figueroa:2020jkf, Cai:2021zsp, Cai:2022erk, Matsubara:2022nbr} and parametric resonance or tachyonic instability \cite{Cai:2018tuh, Chen:2019zza, Chen:2020uhe, Zhou:2020kkf, Peng:2021zon, Addazi:2022ukh, Ashoorioon:2019xqc}. Those enhanced small-scale scalar perturbations also induce sizable GWs via nonlinear couplings, which can exceed the sensitivities of several upcoming GW observations, such as LISA \cite{LISA:2017pwj}, DECIGO \cite{Kawamura:2011zz}, Taiji \cite{Ruan:2018tsw} and TianQin \cite{TianQin:2015yph}. 

We often consider the evolution or generation of tensor fluctuations in the classical field theory with the stochastic initial conditions set by inflation.
As a result, we predict the power spectrum of GWs that is related to the observables such as the CMB power spectrum or GW energy density.
Previous secondary GW studies mainly focus on the auto-power spectrum of second-order scalar-induced gravitational waves (SIGWs), which is a part of the classical stochastic one-loop correction to the primordial tensor power spectrum.
Subleading-order SIGWs, i.e., two-loop corrections, are also investigated in the previous literature \cite{Yuan:2019udt, Zhou:2021vcw, Chang:2022dhh}, which should be subdominant as far as the perturbative expansion is convergent.
Another possible extension at one-loop order is to include the linear vector and tensor perturbations in the second-order source.
References~\cite{Gong:2019mui, Chang:2022vlv} considered the auto-spectrum of the second-order tensor modes sourced by scalar and tensor perturbations.
This class of secondary GWs should also be subdominant unless the internal tensor propagator is more enhanced than the scalar propagators at some scales.
So far, so good. Is there any other source of the secondary GWs?

This paper points out several missing one-loop contributions in previous considerations about SIGWs, i.e., the cross-power spectrum of the first- and third-order tensor fluctuations.
The tensor fluctuation induced by the source with two scalar and one tensor perturbations is third-order in cosmological perturbations, i.e., a subdominant component at the field level. 
However, the cross-power spectrum of the first- and third-order tensor fluctuations is also one-loop whose order in the perturbative expansion is equivalent to that of the induced power spectrum.
Indeed, the iterative solutions and the higher-order interactions are consistently considered in the theory of large-scale structure, where we consider similar classical stochastic loop calculations~\cite{Bernardeau:2001qr}.
Then, there is no reason to ignore those effects in studies of GWs.
Interestingly, Ref.~\cite{Zhou:2021vcw} has already included the iterative solutions for two-loop calculations of induced GWs.
However, they only considered the scalar fluctuations for initial conditions, so the one-loop correction from the cross term of first- and third-order was absent.

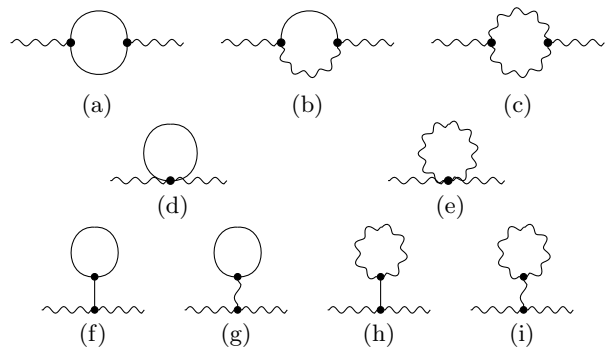
\begin{figure}[htbp]
	\centering
	\begin{minipage}[t]{0.15\textwidth}
		\centerline{
			\begin{tikzpicture}[scale=0.5, transform shape]
			\begin{feynman}
			\vertex (a) ;
			\tikzfeynmanset{every vertex={dot,minimum size=2mm}}
			\vertex [right=of a] (f1);
			\vertex [right=of f1] (f2) ;
			\tikzfeynmanset{every vertex={dot,minimum size=0mm}}
			\vertex [right=of f2] (d);
			\diagram* {
				(a) -- [photon] (f1) [dot] -- [plain, half left,looseness=1.7] (f2)
				-- [plain, half left,looseness=1.7] (f1), 
				(f2) --[photon] (d)
			};
			\end{feynman}
			\end{tikzpicture}}
		\centerline{~(a)}
	\end{minipage}
	\begin{minipage}[t]{0.15\textwidth}
		\centerline{\label{loop222}
			\begin{tikzpicture}[scale=0.5, transform shape]
			\begin{feynman}
			\vertex (a) ;
			\tikzfeynmanset{every vertex={dot,minimum size=2mm}}
			\vertex [dot,right=of a] (f1);
			\vertex [right=of f1] (f2) ;
			\tikzfeynmanset{every vertex={dot,minimum size=0mm}}
			\vertex [right=of f2] (d);
			\diagram* {
				(a) -- [photon] (f1)
				-- [plain, half left,looseness=1.7] (f2)
				-- [photon, half left,looseness=1.7] (f1), 
				(f2) --[photon] (d)
			};
			\end{feynman}
			\end{tikzpicture}}
		\centerline{~(b)}
	\end{minipage}
	\begin{minipage}[t]{0.15\textwidth}
		\centerline{\label{loop222}
			\begin{tikzpicture}[scale=0.5, transform shape]
			\begin{feynman}
			\vertex (a) ;
			\tikzfeynmanset{every vertex={dot,minimum size=2mm}}
			\vertex [dot,right=of a] (f1);
			\vertex [right=of f1] (f2) ;
			\tikzfeynmanset{every vertex={dot,minimum size=0mm}}
			\vertex [right=of f2] (d);
			\diagram* {
				(a) -- [photon] (f1)
				-- [photon, half left,looseness=1.7] (f2)
				-- [photon, half left,looseness=1.7] (f1), 
				(f2) --[photon] (d)
			};
			\end{feynman}
			\end{tikzpicture}}
		\centerline{~(c)}
	\end{minipage}
	
	\begin{minipage}[t]{0.2\textwidth}
		\centerline{
			\begin{tikzpicture}[scale=0.5, transform shape]
			\begin{feynman}
			\vertex (a) ;
			\tikzfeynmanset{every vertex={dot,minimum size=2mm}}
			\vertex [right=of a] (f1);
			\tikzfeynmanset{every vertex={dot,minimum size=0mm}}
			\vertex [above=of f1] (f2);
			\vertex [right=of f1] (b);
			\diagram* {
				(a) -- [photon] (f1) 
				-- [plain, half left,looseness=1.5] (f2)
				-- [plain, half left,looseness=1.5] (f1) , 
				(f1) --[photon] (b)
			};
			\end{feynman}
			\end{tikzpicture}}
		\centerline{~~(d)}
	\end{minipage}
	\begin{minipage}[t]{0.2\textwidth}
		\centerline{
			\begin{tikzpicture}[scale=0.5, transform shape]
			\begin{feynman}
			\vertex (a) ;
			\tikzfeynmanset{every vertex={dot,minimum size=2mm}}
			\vertex [right=of a] (f1);
			\tikzfeynmanset{every vertex={dot,minimum size=0mm}}
			\vertex [above=of f1] (f2);
			\vertex [right=of f1] (b);
			\diagram* {
				(a) -- [photon] (f1) 
				-- [photon, half left,looseness=1.5] (f2)
				-- [photon, half left,looseness=1.5] (f1) , 
				(f1) --[photon] (b)
			};
			\end{feynman}
			\end{tikzpicture}}
		\centerline{~~(e)}
	\end{minipage}
	
	\hspace{3in}
	
	\begin{minipage}[t]{0.1\textwidth}
		\centerline{
			\begin{tikzpicture}[scale=0.44, transform shape]
			\begin{feynman}
			\vertex (a) ;
			\tikzfeynmanset{every vertex={dot,minimum size=2mm}}
			\vertex [right=of a] (f1);
			\vertex [above=1cm of f1] (f2);
			\tikzfeynmanset{every vertex={dot,minimum size=0mm}}
			\vertex [above=of f2] (f3);
			\vertex [right=of f1] (b);
			\diagram* {
				(a) -- [photon] (f1) --[plain] (f2)
				-- [plain, half left,looseness=1.5] (f3)
				-- [plain, half left,looseness=1.5] (f2),
				(f1) --[photon] (b)
			};
			\end{feynman}
			\end{tikzpicture}}
		\centerline{~(f)}
	\end{minipage}
	\begin{minipage}[t]{0.1\textwidth}
		\centerline{\label{loop222}
			\begin{tikzpicture}[scale=0.44, transform shape]
			\begin{feynman}
			\vertex (a) ;
			\tikzfeynmanset{every vertex={dot,minimum size=2mm}}
			\vertex [right=of a] (f1);
			\vertex [above=1cm of f1] (f2);
			\tikzfeynmanset{every vertex={dot,minimum size=0mm}}
			\vertex [above=of f2] (f3);
			\vertex [right=of f1] (b);
			\diagram* {
				(a) -- [photon] (f1) --[photon] (f2)
				-- [plain, half left,looseness=1.5] (f3)
				-- [plain, half left,looseness=1.5] (f2),
				(f1) --[photon] (b)
			};
			\end{feynman}
			\end{tikzpicture}}
		\centerline{~(g)}
	\end{minipage}
	\begin{minipage}[t]{0.1\textwidth}
		\centerline{\label{loop222}
			\begin{tikzpicture}[scale=0.44, transform shape]
			\begin{feynman}
			\vertex (a) ;
			\tikzfeynmanset{every vertex={dot,minimum size=2mm}}
			\vertex [right=of a] (f1);
			\vertex [above=1cm of f1] (f2);
			\tikzfeynmanset{every vertex={dot,minimum size=0mm}}
			\vertex [above=of f2] (f3);
			\vertex [right=of f1] (b);
			\diagram* {
				(a) -- [photon] (f1) --[plain] (f2)
				-- [photon, half left,looseness=1.5] (f3)
				-- [photon, half left,looseness=1.5] (f2),
				(f1) --[photon] (b)
			};
			\end{feynman}
			\end{tikzpicture}}
		\centerline{~(h)}
	\end{minipage}
	\begin{minipage}[t]{0.1\textwidth}
		\centerline{
			\begin{tikzpicture}[scale=0.44, transform shape]
			\begin{feynman}
			\vertex (a) ;
			\tikzfeynmanset{every vertex={dot,minimum size=2mm}}
			\vertex [right=of a] (f1);
			\vertex [above=1cm of f1] (f2);
			\tikzfeynmanset{every vertex={dot,minimum size=0mm}}
			\vertex [above=of f2] (f3);
			\vertex [right=of f1] (b);
			\diagram* {
				(a) -- [photon] (f1) --[photon] (f2)
				-- [photon, half left,looseness=1.5] (f3)
				-- [photon, half left,looseness=1.5] (f2),
				(f1) --[photon] (b)
			};
			\end{feynman}
			\end{tikzpicture}}
		\centerline{~(i)}
	\end{minipage}
	\caption{\label{threeloop}
		Nine typical one-loop contributions to the tensor power spectrum. The internal wavy and solid lines represent tensor and scalar propagators, respectively. 
		The standard second-order SIGWs $\mathcal{P}_h^{(22)}$ is included in diagram (a). 
		Diagrams (b) and (c) were studied in Ref. \cite{Gong:2019mui}, which are subdominant unless the tensor propagator is more enhanced than the scalar one. (d) is the new graph considered in this paper. We ignore (e, h, i) since we only focus on the enhancement of the scalar propagator. The tadpole (g) 	is zero, and (f)  is also negligible in the IR region. Note that we ignored vectors for simplicity.}
\end{figure}

\begin{figure}[htp]
	\centering
	\includegraphics[width=\linewidth]{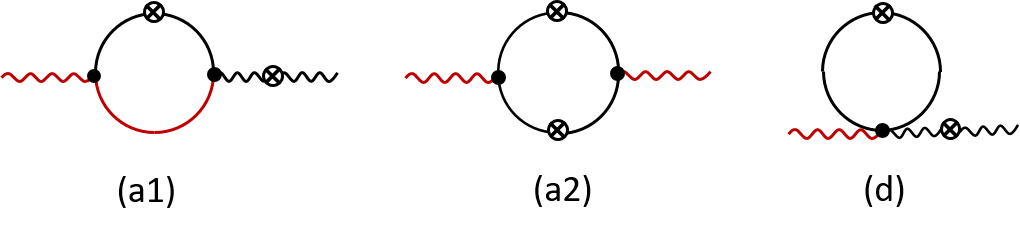}
	\caption{Detailed structures of diagrams (a) and (d) shown in Fig.~\ref{threeloop} . 
	We draw the diagrams following Ref.~\cite{Crocce:2005xy}, while the direction of time is implicit for simplicity.
	The cross circle represents the contraction between two linear fields (i.e., their power spectra), which is also regarded as an ``external source'' in this system.
			The red lines are Green functions. 
			The diagrams (a1) and (d) contribute to $\mathcal{P}^{(13)}$, while (a2) is the induced GWs $\mathcal{P}^{(22)}$.}
	\label{fig:detail}
\end{figure}

We consider all possible sources up to third-order~(see Tab.~\ref{table1}, and also Figs.~\ref{threeloop} and \ref{fig:detail}).
We will show that the IR behavior of the new correction is very different from that of the auto-spectrum discussed in the previous literature and can be important for the actual gravitational wave measurements.
A recent work \cite{Ota:2022hvh} also reported the one-loop quantum corrections to PGWs by an excited spectator field, using the in-in formalism during inflation.
Their consistent loop calculation showed that superhorizon PGWs are amplified or suppressed by the loop effect.  
Inspiringly, these astonishing results show the possibility of probing the extremely small-scale phenomena during inflation with large-scale GW observations.
In this paper, we consider the classical counterpart of their scale-invariant corrections.
We will show a similar effect in a classical setup in universes dominated by radiation or dust.

\begin{table}
\caption{\label{table1}
Possible source terms at third order, including tensor and scalar perturbations. 
The first line corresponds to the Born approximation, and the first iterative source is shown in the second to the fifth line.
$h$ and $\phi$ represent the tensor and scalar perturbations, respectively. $\phi$ implies either the curvature perturbation $\Psi$ or gravitational potential $\Phi$.
A product of $\phi$ and $h$ in a subscript implies the source of the corresponding second-order perturbations (e.g. $h_{\phi h}$ means $h^{(2)}$ sourced by $h^{(1)}\phi^{(1)}$).
The underlined terms are correlated with linear tensor modes.
(a) to (i) indicate the corresponding diagrams in Fig.~\ref{threeloop}~(note that Green functions are implicit.).
}
\begin{ruledtabular}
\begin{tabular}{lcccc}

Born approx.& $\phi\phi\phi$ & $^{(\rm d)}\underline{h \phi\phi}$    & $hh\phi$  & $^{(\rm e)} \underline{hhh}$  \\
1st iteration &  $\phi h_{\phi \phi}$ &  $^{(\rm b)}\underline{\phi h_{\phi h}}$    &  $\phi h_{h h}  $  &   \\
&  $^{(\rm f)}\underline{h \phi_{\phi \phi}}$ &  $h \phi_{\phi h}$    &  $^{(\rm b,h)} \underline{h \phi_{h h} } $  &   \\
& $\phi\phi_{\phi \phi}$ &  $^{(\rm a)}\underline{\phi\phi_{\phi h}}$    &  $\phi\phi_{h h}  $ &   \\
&  $^{(\rm g)}\underline{hh_{\phi \phi}}$ &  $hh_{\phi h}$    &  $^{(\rm c,i)}\underline{hh_{h h}}  $    &
\end{tabular}
\end{ruledtabular}
\end{table}

The paper is organized as follows. In Sec. \ref{Basic}, we extend 
cosmological perturbation theory to 
third-order by including the linear tensor fluctuations to the nonlinear source. Then, we derive a generic form for the missing one-loop correction to secondary GWs, i.e., the cross-power spectrum $\mathcal{P}_h^{(13)}$. 
In Sec. \ref{Sec:IR}, we calculate $\mathcal{P}_h^{(13)}$ in the cases of both radiation-dominated (RD) and matter-dominated (MD) eras with a delta-function-like scalar source. Their IR behaviors are also investigated in detail. 
In Sec. \ref{exp}, we elaborate on the influence on the tensor-to-scalar ratio from $\mathcal{P}_h^{(13)}$ in terms of the collaborative multi-frequency GW experiments for PBH detection.
Finally, we summarize the results in Sec.  \ref{conclusion}.

\section{Basic equations and sources for classical one-loop correction} 
\label{Basic}

In this section, we first derive the evolution equation for tensor fluctuations in the presence of two scalars and one tensor.
Then we integrate the equation of motion using the Green function method.
In the standard SIGW calculation, one only expands the source to the second order in scalar perturbations, but we go beyond the expansion, including additional first-order tensor fluctuations.

First, we perturb the metric tensor around the spatially flat Friedmann–Lema\^{\i}tre–Robertson–Walker metric to the nonlinear order in scalar perturbations and tensor fluctuations. 
In the conformal Newtonian gauge, the metric takes the form of
\begin{align}\label{metric}
    \mathrm{d}s^2=-a^2(1+2\Phi) d\tau^2+a^2(1-2\Psi)(\delta_{ij}+h_{ij})dx^i dx^j,
\end{align}
and the tensor fluctuation is expanded into
\begin{align}
	h_{ij} &= h_{ij}^{(1)}+\frac{1}{2}h_{ij}^{(2)}+\frac{1}{6}h_{ij}^{(3)}+\cdots,
\end{align}
where the superscript in parentheses implies the order in cosmological perturbations, and we dropped vector perturbations for simplicity.
Indeed, this paper considers a delta-function-like sharp peak for scalar initial conditions, which should not induce the second-order vector perturbations from momentum conservation. So, we ignore the vector perturbations up to second order safely~\cite{Ota:2020vfn,Chang:2022dhh}.
$h^{(1)}_{ij}$ is the linear tensor mode initially set by inflation, $h^{(2)}_{ij}$ is the induced tensor mode discussed in the previous literature, and $h^{(3)}_{ij}$ is the new contribution sourced at third order in cosmological perturbations.
Note that $h^{(3)}_{ij}$ from the scalar fluctuations was considered in Ref.~\cite{Zhou:2021vcw} for two-loop-order auto-power spectrum of $h^{(3)}_{ij}$.
The transverse-traceless condition is not unique when expanding the metric tensor to nonlinear orders~\cite{Maldacena:2002vr}.
We impose the conditions for $h^{(n)}_{ij}$ in Eq.~\eqref{metric}, i.e., $\partial^i h^{(n)}_{ij}=h^{(n)i}{}_{i}=0$, where the Latin indices are raised and lowered by the background spatial metric $\delta^{ij}$ and $\delta_{ij}$.
The Fourier integral of the tensor perturbation is written as
\begin{align}
    h^{(n)}_{ij}(\tau,\boldsymbol{x})=\int\frac{d^3k}{(2\pi)^3}\sum_{\lambda=+,\times}e^{\lambda}_{ij}(\hat{k})h^{(n)\lambda}_{\boldsymbol{k}}(\tau)e^{i\boldsymbol{k\cdot x}},
\end{align}
with two orthonormal polarization bases defined as 
\begin{align}
    &e^{(+)}_{ij}(\hat{k})=\frac{1}{\sqrt{2}}\left[ e_i(\hat{k})e_j(\hat{k})-\overline{e}_i(\hat{k})\overline{e}_j(\hat{k})\right],\\
   &e^{(\times)}_{ij}(\hat{k})=\frac{1}{\sqrt{2}}\left[ e_i(\hat{k})\overline{e}_j(\hat{k})+\overline{e}_i(\hat{k})e_j(\hat{k})\right],
\end{align}
where $e_i(\hat{k})$ and $\overline{e}_i(\hat{k})$ are a set of orthonormal vectors perpendicular to $\boldsymbol{k}$, and $\hat{k}\equiv\boldsymbol{k}/|\boldsymbol{k}|$.
We define the dimensionless power spectrum of the tensor fluctuations as
\begin{align}
    \langle h^{(n)\lambda}_{\boldsymbol{k}}h^{(m)\lambda'}_{\boldsymbol{k}'}\rangle=\delta_{\lambda\lambda'}(2\pi)^3\delta^3(\boldsymbol{k+k'})\frac{2\pi^2}{k^3}\mathcal{P}^{(nm)}_{h,\lambda}.
\end{align}
Hereafter we omit the polarization index when we do not have to specify a polarization component. 

We go beyond the Born approximation in this paper, so the scalar fluctuations should be included up to second order:
\begin{align}
		\Phi &= \Phi^{(1)} + \frac{1}{2}\Phi^{(2)},
		\\
		\Psi &= \Psi^{(1)} + \frac{1}{2}\Psi^{(2)}.\label{phipsi}
\end{align}
$\Phi^{(2)}$ and $\Psi^{(2)}$ are the first iterative corrections relevant to the one-loop order contribution in the end.
These terms contribute to the diagrams (a, b, f, h)  in Fig.~\ref{threeloop}.
The same contributions were also considered in Ref.~\cite{Ota:2022hvh} in the context of the one-loop inflationary power spectrum.
We assume that anisotropic stress is negligible at linear order, so the gravitational potential and curvature perturbation are equivalent.
Hereafter, we denote 
\begin{align}
	\Phi^{(1)}=\Psi^{(1)}=\phi,	
\end{align}
for notational simplicity.

Expanding the Einstein equation and projecting it onto the polarization plane, one finds
\begin{align} \label{eom_h}
    h_{ij}^{(n)}{}''+2\mathcal{H}h^{(n)}_{ij}{}'-\nabla^2h_{ij}^{(n)}=\mathcal{T}_{ij}^{lm}S^{(n)}_{lm},
\end{align}
where $\mathcal{H}\equiv a'/a=aH$ is the comoving Hubble parameter, and the prime denotes the derivative with respect to the conformal time $\tau$. 
$\mathcal T^{lm}_{ij}$ is the projection operator onto the transverse-traceless plane.
In Fourier space, we recast Eq.~\eqref{eom_h} into 
\begin{align} \label{eom_h_k}
    h^{(n)}_{\boldsymbol{k}}{}'' +2\mathcal{H}h^{(n)}_{\boldsymbol{k}}{}'+k^2 h_{\boldsymbol{k}}^{(n)}=S^{(n)}_{\boldsymbol{k}},
\end{align}
where we define
\begin{align}
	S^{(n)}_{\boldsymbol{k}}\equiv e^{ij}(\hat{k}) S^{(n)}_{ij,{\bm k}}.
\end{align}
Given a source term, one can integrate Eq.~\eqref{eom_h_k} using the Green function:
\begin{align}
    h^{(n)}_{\bm k}(x)=\int^x d\Tilde{x}\frac{a(\Tilde{x})}{a(x)} kG^h_{\bm{k}}(x,\Tilde{x})\frac{S^{(n)}_{\bm k}(\Tilde{x})}{k^2},
\end{align}
where $x\equiv k\tau$.
$G^h_{\bm k}$ is the Green function for tensor modes which takes the following form in RD universe
\begin{align}
kG^h_{\bm{k}, {\rm RD}}(x,\Tilde{x})=
\sin\left(x-\Tilde{x}\right),
\end{align}
where $k\equiv|{\bm k}|$. In the MD universe, we have
\begin{align}
kG^h_{\bm{k}, {\rm MD}}(x,\Tilde{x})=
x\Tilde{x}\left[y_1(x)j_1(\Tilde{x})-j_1(x)y_1(\Tilde{x})\right],
\end{align}
where $j_1(x)$ and $y_1(x)$ are the spherical Bessel functions of the first and second kind, respectively.

Possible terms in $S^{(3)}_{lm}$ are summarized in Tab.~\ref{table1}, and terms correlated with first-order tensor fluctuations are underlined.
We will consider the cross-power spectrum between the first- and third-order tensor modes so we do not have to evaluate the most general forms.
The operator products that appear in the final spectrum are, up to the transfer functions and Green functions, written as  
\begin{align}
	\langle h^{(1)}_{\boldsymbol{k'}}h^{(3)}_{\boldsymbol{k}}\rangle' \sim \int \frac{d^3\mathbf{p}d^3\mathbf{q}}{(2\pi)^6}  \langle h_{\boldsymbol{k'}}h_{\boldsymbol{k-p-q}}\phi_{\boldsymbol{p}}\phi_{\boldsymbol{q}}\rangle',
	\label{simp}
\end{align}
where the prime on the bracket implies that we drop a delta function with respect to the external momenta. 
The RHS of Eq.~\eqref{simp} reduces to 
\begin{align}
	\langle h_{\bm k'}h_{\bm k} \rangle' \int \frac{d^3\mathbf{p}}{(2\pi)^3}  \langle\phi_{\bm p}\phi_{-\bm p}\rangle'\label{h1h3}.
\end{align}
Thus, tensor fluctuation decouples from the loop, and the integral structure is simplified.
Therefore, we justify the following replacement in the source:
\begin{align}
	\int \frac{d^3\mathbf{p}d^3\mathbf{q}}{(2\pi)^6} h_{\boldsymbol{k-p-q}}\phi_{\boldsymbol{p}}\phi_{\boldsymbol{q}} \to h_{\bm k} \int \frac{d^3\mathbf{p}}{(2\pi)^3}  \phi_{\bm p}\phi_{-\bm p}.\label{premise}
\end{align}
Finally, cross-correlating the third-order tensor fluctuation with the linear one, we obtain
\begin{align} \label{P13}
    \mathcal{P}^{(13)}_h(\tau,k)=\mathcal{P}^{(11)}_h(k) \int\frac{du}{u}I_h(u,x) \mathcal{P}^{(11)}_\phi(ku),
\end{align}
where we use Eq.~\eqref{premise} and defined $u\equiv p/k$. $\mathcal{P}^{(11)}_h$ and $\mathcal{P}^{(11)}_\phi$ are the initial linear power spectra of tensor and scalar perturbations, respectively.
All details about the source are included in the kernel function
\begin{align} \label{kernel}
    I_h(u,x)=T_h(x)\int^{x}d\Tilde{x} \frac{a(\Tilde{x})}{a(x)}kG^h_{\boldsymbol{k}}(x,\Tilde{x})f_h(u,\Tilde{x}),
\end{align}
where $T_h$ is the linear transfer function for tensor fluctuations, and $f_h(u,x)$ can be found from the concrete calculation of the source function discussed below.

\subsection{Born approximation}
The first relevant correction is the Born approximation for the third-order source.
Dropping other irrelevant terms, the third-order source from the triple product of the pure first-order perturbations is given as
\begin{align}
&S^{(3)}_{h\phi\phi,ij}\equiv h^{(1)}_{ij}\Big[16\phi\nabla^2\phi+\frac{8 (1+3\omega)}{3(1+\omega)}(\nabla \phi)^2-\frac{32 \nabla\phi \nabla\phi'}{3\mathcal{H}(1+\omega)}\nonumber\\
&-\frac{16\left(\nabla \phi'\right)^2}{3\mathcal{H}^2(1+\omega)}\Big]-24\phi\Big(3\phi'h^{(1)}_{ij}{}'+2\phi\nabla^2h^{(1)}_{ij}\Big)\nonumber\\
&-\frac{16\partial^k \phi}{\mathcal{H}^2(1+\omega)}\Big[\left(\mathcal{H}\partial_j\phi+\partial_j\phi'\right)h^{(1)}_{k  i}{}'+\left(\mathcal{H}\partial_i\phi+\partial_i\phi'\right)h^{(1)}_{k  j}{}'\Big]\nonumber\\
&+24\phi\partial^k \phi\Big(\partial_j h^{(1)}_{k  i}+\partial_i h^{(1)}_{k  j}-\partial_k  h^{(1)}_{ij}\Big),
\end{align}
where we simplified the source, using the equations of motion for $\phi$ and $h^{(1)}_{ij}$.
Under the premise of Eq.~\eqref{premise}, we find 
\begin{align}\label{source1}
    S^{(3)}_{h\phi\phi,\bm k}
    &=-\int\frac{d^3\mathbf{p}}{(2\pi)^3} \bigg[ \frac{8(5+3\omega)}{3(1+\omega)}p^2h_{\bm k}\phi_{\bm p}\phi_{-\bm p}
    \nonumber 
    \\
    & +\frac{32}{3(1+\omega)\mathcal{H}} p^2 \big( h_{\bm k}\phi'_{\bm p}\phi_{-\bm p} + h'_{\bm k}\phi_{\bm p}\phi_{-\bm p} \big)  \nonumber
    \\&
    +\frac{16}{3(1+\omega)\mathcal{H}^2} p^2 \big( h_{\bm k}\phi'_{\bm p}\phi'_{-\bm p} + 2 h'_{\bm k}\phi'_{\bm p}\phi_{-\bm p} \big) \nonumber
   \\
    &+72h_{\bm k}'\phi_{\bm p}'\phi_{-\bm p}-48k^2h_{\bm k}\phi_{\bm p}\phi_{-\bm p} \bigg].
\end{align}  
Then we get
\begin{align}
&f_{h, h\phi\phi}(u,x)
\notag \\
\equiv &- \Big\{\frac{8(5+3\omega)}{3(1+\omega)}u^2T_h(x)T_\phi(ux)T_\phi(ux)
    \notag \\
    &\left.+\frac{8(1+3\omega)^2}{3(1+\omega)}u^2x^2T'_h(x)T'_\phi(ux)T_\phi(ux)\right.\nonumber\\
     &\left.-48T_h(x)T_\phi(ux)T_\phi(ux)+72T'_h(x)T'_\phi(ux)T_\phi(ux)
    \right. \notag \\
     &\left.+\frac{4(1+3\omega)^2}{3(1+\omega)}u^2x^2T_h(x)T_\phi'(ux)T_\phi'(ux)\right.\nonumber\\
     &\left.+\frac{16(1+3\omega)}{3(1+\omega)}u^2x\left[T_h(x)T_\phi'(ux)T_\phi(ux)
     \right.\right.
     \notag \\
     &\left.+T_h'(x)T_\phi(ux)T_\phi(ux)\right] \Big\},
     \label{sourcef1}
\end{align}
where we used the conformal Hubble parameter written as
\begin{align}
	\mathcal{H} = \frac{2}{(1+3\omega)\tau}.
\end{align}
Note that the prime denotes the derivative with respect to $x$ hereafter. $T_h$ and $T_\phi$ are transfer functions for the linear tensor and scalar modes, respectively. 
Following the similar treatments in Ref.~\cite{Kohri:2018awv}, one can find the analytical solution of the kernel function $I_{h,h\phi\phi}(u,x)$.

\subsection{First iterative solution}

In addition to the Born approximation, we need to account for the first iterative solution for the one-loop-order correction to the cross-power spectrum.
We summarized possible sources in Tab.~\ref{table1}, but it is found that only $\phi \phi_{\phi h}$ is the relevant contribution in our case.
$\phi h_{\phi h }$, $h h_{\phi \phi}$ and $h \phi_{\phi \phi}$ are potentially comparable to $\phi \phi_{\phi h}$.
However, these terms contribute in the subhorizon region where the induced GWs dominate, as discussed in Appendix~\ref{app3}.
As discussed in the next section, we may ignore those contributions when we are interested in the IR region.
We also safely ignore $hhh$, $hh_{hh}$, $h \phi_{h h}$ since they are not amplified by scalar fluctuations.

The source terms from $\phi$ and second-order scalar induced by $\phi$ and $h$ are written as
\begin{align} \label{Sphiphih}
&S^{(3)}_{\phi\phi_{\phi h},ij} \nonumber
\\&=
{8 \over \mathcal{H}^2(1+\omega)}
\Big[
\big( \mathcal{H}\partial_i\phi + \partial_i\phi' \big)
\big( \mathcal{H}\partial_j\Phi^{(2)} + \partial_j\Psi^{(2)}{}' \big)
+\big(i\leftrightarrow j\big)
\Big] \nonumber
\\
&-12\Big[
\big(\Phi^{(2)} + \Psi^{(2)}\big) \partial_i\partial_j\phi + \phi \partial_i\partial_j\big(\Phi^{(2)} + \Psi^{(2)}\big) \nonumber
\\&
+ \partial_i\phi\partial_j\Psi^{(2)}
+ \partial_j\phi\partial_i\Psi^{(2)}
\Big] ~.
\end{align}

The difference of $\Phi^{(2)}$ and $\Psi^{(2)}$ arises from $\mathcal{O}(\phi^2,\phi\partial h)$.
Hence, we can take $\Phi^{(2)}=\Psi^{(2)}$ and drop derivative terms like $\partial h$ in the following calculation for the IR region.
The equation of motion for $\Psi^{(2)}$ is given as
\begin{align}\label{eomphi2}
    \Psi^{(2)}{}''+3\mathcal{H}(1+\omega)\Psi^{(2)}{}'-\omega\nabla^2\Psi^{(2)}=-2\omega h^{(1)}_{ij}\partial_{i}\partial_j\phi .
\end{align}
We immediately find $\Psi^{(2)}=0$ for the matter era where $\omega =0$.
During the radiation era, we get~\cite{Inomata:2020cck}
\begin{align}
    \Psi^{(2)}_{\bm{k}+\bm{p}}(\tau)=\frac{1}{\sqrt{2}}\sin^2\theta\cos2\varphi I_{\phi,\text{\tiny{RD}}}(u,v,x)\phi^0_{\bm{p}}h^0_{\bm{k}},
    \label{psi2}
\end{align}
where $h^0_k$ and $\phi^0_p$ 
are the initial value at the superhorizon scale, and
\begin{align}
    I_{\phi,\text{\tiny{RD}}}(u,v,x)&=\int^xd\Tilde{x}\frac{\Tilde{x}^2}{x^2}kG^s_{\bm{k}}(vx,v\Tilde{x})\frac{u^2}{v}\frac{2}{3} T_h(\Tilde{x})T_\phi(u\Tilde{x}).
\end{align}
Note that we defined $v\equiv|\bm{p}+\bm{k}|/k$, and $(\theta,\varphi)$ are the spherical coordinates of $\bm{p}$ with respect to $\hat{z}\parallel\bm{k}$, and the Green function for scalar perturbations is
\begin{equation}
kG^s_{\bm{k}}(x,\Tilde{x}) =\frac{x\Tilde{x}}{\sqrt{3}}\left[j_1\left(\frac{\Tilde{x}}{\sqrt{3}}\right)y_1\left(\frac{x}{\sqrt{3}}\right)-\big(x\leftrightarrow\Tilde{x}\big)\right].
\end{equation}
Substituting Eq.~\eqref{psi2} into Eq.~\eqref{Sphiphih}, we find the first iterative solution.
The rest of the calculation is the same as the Born approximation, so we find
\begin{align}\label{sourcef2}
&f_{h,\phi\phi_{\phi h}}(u,v,x)= \int d\theta\sin^5\theta\,\frac{3}{2}u^2\nonumber
\\
&\times\big[3T_\phi I_{\phi,\text{\tiny{RD}}}+x\left(T'_\phi I_{\phi,\text{\tiny{RD}}}+T_\phi I'_{\phi,\text{\tiny{RD}}}\right)+x^2T'_\phi I'_{\phi,\text{\tiny{RD}}}\big].
\end{align}

\begin{figure*}[htp]
    \centering
    \includegraphics[width=\linewidth]{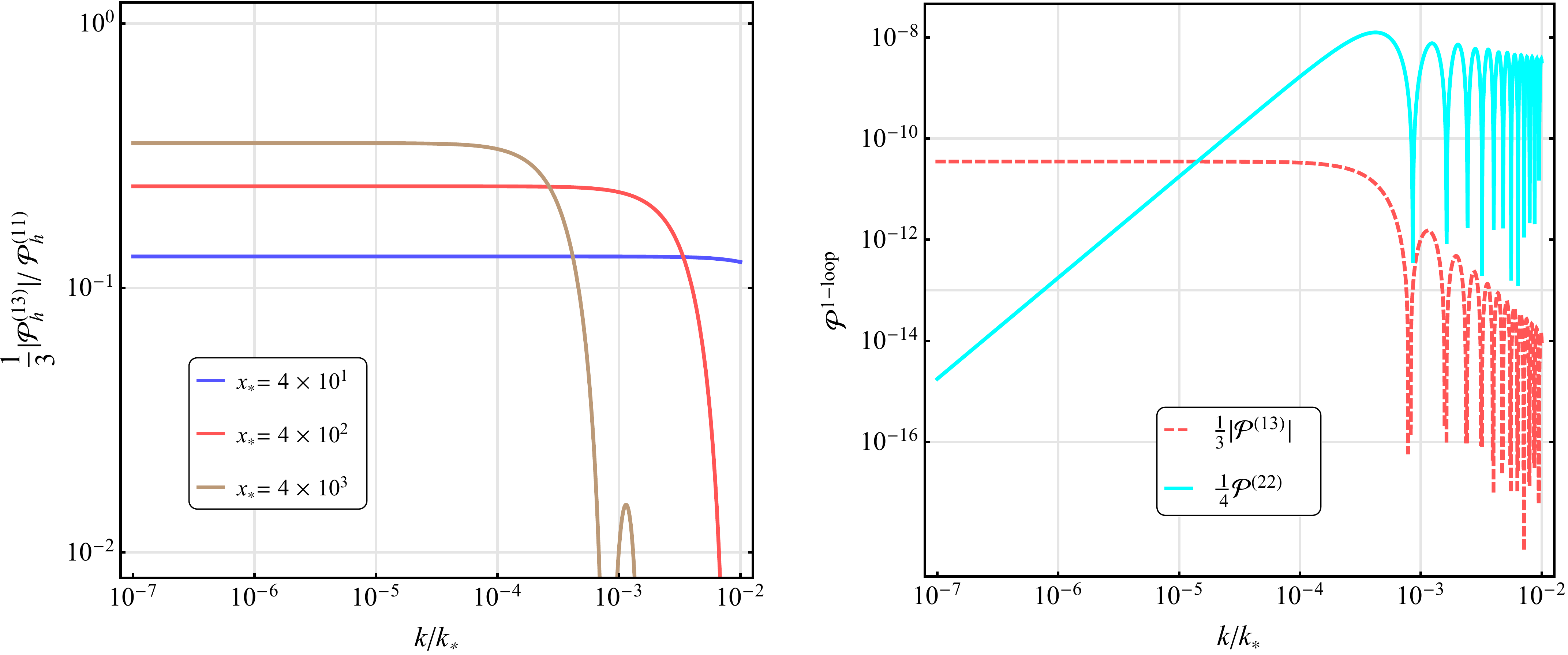}
    \caption{\textit{Left}: the absolute value of the one-loop correction $\frac{1}{3}\big|\mathcal{P}^\text{(13)}_h\big|$ shown in Eq. \eqref{P13x} as a function of $k/k_*$ during the RD era. The vertical axis is normalized by the primordial tensor spectrum $\mathcal{P}^{(11)}_h$. Brown, red and blue solid lines denote different times $y\equiv k_* \tau=4\times10^1, 4\times10^2$ and $4\times10^3$, respectively. The damping feature for $k/k_*\gtrsim y^{-1}$ implies the horizon entry.
    {\it Right:} a comparison between the amplitudes of the one-loop corrections $\big|\mathcal{P}_h^{(13)}\big|$ and the standard second-order SIGWs $\mathcal{P}_h^{(22)}$ during the RD era, which are shown by the red dashed and the cyan solid curves, respectively. The parameters 
    are taken as $y_{\rm D} =4\times 10^3$, $A_\zeta=10^{-2}$, and $\mathcal{P}^{(11)}_h=10^{-10}$.}
    \label{RD}
\end{figure*}

\section{IR behavior of the one-loop correction}
\label{Sec:IR}

In the previous section, we obtained the new one-loop contribution 
\begin{align} \label{P13x}
    \mathcal{P}^{(13)}_h(\tau,k)=\mathcal{P}^{(11)}_h(k) \int\frac{du}{u}\left(I_{h,h\phi\phi}+I_{h,\phi\phi_{\phi h}}\right) \mathcal{P}^{(11)}_\phi(ku),
\end{align}
where the kernels $I_{h,h\phi\phi}$ and $I_{h,\phi\phi_{\phi h}}$ are defined through Eq.~(\ref{kernel}), with the source integrals $f_{h,h\phi\phi}$ and $f_{h,\phi\phi_{\phi h}}$ given by Eqs.~(\ref{sourcef1}) and (\ref{sourcef2}), respectively.

Equation~\eqref{P13x} implies that the new one-loop contributions are roughly written as $\mathcal P^{(13)}_h\sim \mathcal P^{(11)}_h\mathcal P^{(11)}_{\phi}$. In comparison, the SIGW auto-power spectrum is $\mathcal P^{(22)}_h\sim (\mathcal P^{(11)}_{\phi})^2$. 
Therefore, one may naively expect that the new contributions are suppressed by a factor of $\mathcal P^{(13)}_h/\mathcal P^{(22)}_h \sim \mathcal P^{(11)}_h/\mathcal P^{(11)}_{\phi}$, which should be the subdominant of secondary GWs. Is that true?
Equation~\eqref{premise} suggests that the new third-order correction is not a production of GWs from zero but a modulation of primordial tensor fluctuations due to couplings between tensor and scalar.
The physical origin differs from the GW production, so the property is not necessarily the same.
Indeed, we will show that the new contribution can be dominant in the IR region.

In this section, we concretely compute $\mathcal P^{(13)}_h$ for a simple but phenomenologically interesting delta-function-like scalar power spectrum.
We often consider the delta-function-like spectrum for the SIGW counterpart of PBH formation scenarios, and the SIGWs from the source have peaks near the sharp peak of scalar fluctuations.
In the IR region, $\mathcal P^{(22)}_h$ is suppressed as SIGWs are causally generated from physical processes. 
Here, we will show that the IR behavior of the new correction is very different from that of the auto-power spectrum of the SIGWs \cite{Cai:2019cdl}, and $\mathcal P^{(13)}_h$ can be dominant for large-scale tensor fluctuations. 
In the following, we consider universes dominated by radiation and dust separately and compare their behaviors with the standard SIGWs.

\subsection{Radiation-dominated era}\label{RDsection}

During the RD era, $\omega = 1/3$, and the transfer functions are given as
\begin{align}
    T_\phi(x)&=\frac{9}{x^2}\left[\frac{\sin (x/\sqrt{3})}{x/\sqrt{3}}-\cos\left(\frac{x}{\sqrt{3}}\right)\right],\label{Tp_RD}\\
    T_h(x)&=j_0(x)=\frac{\sin{x}}{x}.\label{Th_RD}
\end{align}
The source functions are the sum of 
\begin{align}
   f_{h,h\phi\phi}&=T_h\left[(48-12u^2)T_\phi T_\phi-8u^2xT_\phi'T_\phi-4u^2x^2T_\phi'T_\phi'\right]\nonumber\\
   &-T'_h\left(8u^2x^2T'_\phi T_\phi+8u^2xT_\phi T_\phi+72T'_\phi T_\phi\right),
   \label{fh1}
   \\
   f_{h,\phi\phi_{\phi h}}&=\int d\theta\sin^5\theta\,\frac{3}{2}u^2
\notag \\
& \times\big[3T_\phi I_{\phi,\text{\tiny{RD}}}+x\left(T'_\phi I_{\phi,\text{\tiny{RD}}}+T_\phi I'_{\phi,\text{\tiny{RD}}}\right)+x^2T'_\phi I'_{\phi,\text{\tiny{RD}}}\big]\label{fh2}.
\end{align}
where the arguments are suppressed for notational simplicity, while $T_\phi$ and $T_h$ are functions of $ux$ and $x$, respectively.
We consider the delta-function-like source amplified at $k=k_*$, i.e.,
\begin{align}\label{deltasource}
    \mathcal{P}^\delta_\zeta(k)=A_\zeta\delta(\ln k- \ln k_*),
\end{align}
where $A_\zeta$ is the overall amplitude. 
The relationship between the super horizon comoving curvature perturbation $\zeta$ and Newtonian potential $\phi$ is
\begin{align}
\phi_{\bm k }=\frac{3+3\omega}{5+3\omega}\zeta_{\bm k}.\label{PphitoPzeta}
\end{align}
Combining Eqs.~\eqref{P13}, \eqref{kernel}, and \eqref{Tp_RD} to \eqref{PphitoPzeta}, we reach the final result of $\mathcal{P}^{(13)}_h(k,\tau)$.
Eq.~(\ref{fh2}) contains two layers of Green function integrals, whose analytical result is tedious.  
However, we can simplify the expressions in the IR region since the solution of Eq.~(\ref{eomphi2}) is written as $\Psi^{(2)}_{\bm p}(\tau)\sim \tau \phi'_{\bm p}(\tau)$ for the superhorizon tensor modes.

The final expression has a factorized form
\begin{align} \label{p13_final}
    \frac{1}{6}\mathcal{P}^{(13)}_h(k, \tau)\equiv\mathcal{F}(k_*, k, \tau) \mathcal{P}^{(11)}_h(k) ,
\end{align} 
where the prefactor $\mathcal{F}$ for $k\tau < 1$ scales as
\begin{align}
    \mathcal{F}(k_*,k,\tau)\bigg|_{k\tau < 1} &\simeq A_\zeta\left[ 2-2.4 \log (k_*\tau)+\mathcal O\left(\frac{k}{k_*}\right)\right].
    \label{Famp}
\end{align}
Thus $\mathcal F$ is $k$-independent for $k/k_*\ll 1$ modes, and the one-loop correction $\mathcal{P}^{(13)}_h$ has the same scaling as the linear power spectrum $\mathcal{P}_h^{(11)}$ on the superhorizon scales.
The above scale dependence arises from Eq.~(\ref{fh1}) and Eq.~(\ref{fh2}), but their signs are opposite.
They are partly canceled by each other, and the total contribution is dominated by Eq.~(\ref{fh1}). 
As a result, the new one-loop-order correction decreases the primordial spectrum. We show the analytical expression from the dominated source $h\phi\phi$ in Appendix \ref{app2}.
Since $\mathcal F$ is linear with respect to $\mathcal P^\delta_\zeta$, we can straightforwardly generalize Eq.~\eqref{Famp} for an arbitrary scalar spectrum.
As we have
\begin{align}
	\mathcal P_\zeta(k) = \int d\ln k_* \mathcal P_\zeta(k_*) \delta(\ln k - \ln k_*),
\end{align}
we find
\begin{align}
	\mathcal{F}(k,\tau) = \frac{1}{A_\zeta} \int d\ln k_*  \mathcal P_\zeta(k_*) \mathcal{F}(k_*,k,\tau).\label{superposition}
\end{align}

The $\tau$-dependence of Eq.~\eqref{Famp} implies that the scalar peak at $k=k_*$ contributes to $\mathcal{P}_h^{(13)}$ at any time after the horizon re-entry of the scalar peak.
This is because we ignored the shear viscosity in the radiation fluid for simplicity.
With this approximation, sound waves propagate forever.
In the real Universe, photons are coupled to electrons via Compton scattering, which introduces acoustic dissipation.
Then the inhomogeneity inside the diffusion scale is smeared.
As a result, the gravitational potential and curvature perturbations are erased.
The photon diffusion scale is given by~\cite{Hu:1995em}
\begin{align}
	k_{\rm D}&\sim 2.34\times 10^{-5} \Theta_{2.7}(1-Y_{\rm p}/2)^{1/2}\Omega_b^{1/2} z^{3/2}h/{\rm Mpc}
	\notag \\
	&\sim 4.9\times 10^{-6} z^{3/2} h/{\rm Mpc},
\end{align}
with a normalized CMB temperature $\Theta_{2.7}=T_{\rm CMB}/2.7=1.01$, the primordial helium mass fraction $Y_{\rm p}=0.23$ and the baryon energy density fraction $\Omega_{\rm b} =0.0486$.
Then, the scalar fluctuations with $k>k_{\rm D}$ are exponentially suppressed because of the diffusion effect.
Solving $k_*=k_{\rm D}(z(\tau_{\rm D}))$ for $\tau_{\rm D}$, we find the final amplitude at $\tau=\tau_{\rm D}$, that is, $\mathcal{F}(k_*,k,\tau_{\rm D})$.
The source vanishes for $\tau>\tau_{\rm D}$, so the superhorizon tensor fluctuations should not vary anymore.
Since the loop momentum in Eq. \eqref{P13} is independent of the external momentum $k$, momentum conservation does not introduce any additional factor. In contrast, the Heaviside step function appears due to momentum conservation in $\mathcal P^{(22)}_h$~\cite{Kohri:2018awv}.

Let us consider a specific case with $k_* \sim 10^5 h /\text{~Mpc}$ as a reference scale for PBH formation during the RD era, corresponding to tens of solar masses which may account for LIGO/Virgo GW detection events \cite{Bird:2016dcv, Sasaki:2016jop}.
The conformal time when $k_*\sim 10^5h/{\rm Mpc}$ enters the diffusion scale is estimated to be $\tau_{\rm D}\sim 0.04{\rm Mpc}/h$. 
Then we find $y_{\rm D} \equiv k_*\tau_{\rm D} \sim 4\times10^3$.
In the left panel of Fig.~\ref{RD}, we plot $\big|\mathcal{P}^{(13)}_h\big|/3$, where the coefficient $1/6$ comes from the metric decomposition \eqref{metric} and a factor of 2 appears in the cross term. 
Brown, red, and blue solid lines in the left panel of Fig.~\ref{RD} denote different times $y(\equiv k_*\tau)=4\times10^1, 4\times10^2$ and $4\times10^3$, respectively, and we normalize the total power spectrum by the linear power spectrum $\mathcal{P}^{(11)}_h$.  
We find that the one-loop correction $\mathcal{P}_h^{(13)}$ results in a {\it negative constant} on the superhorizon scales, which suppresses the scale-invariant primordial tensor power spectrum. 
With observationally
allowable curvature perturbations~\cite{Inomata:2018epa, Green:2020jor} (also see Fig.~\ref{fig:observation}), $A_\zeta < 10^{-2}$ at $k_* \sim 10^5 h \text{~Mpc}^{-1}$, we find that the initial tensor power spectrum loses the amplitude by at most 35\%. 
After the horizon re-entry, the tensor modes evolve as if they are the linear tensor modes described by the red dashed curve in the right panel of Fig.~\ref{RD}.
We emphasize that $\mathcal{P}^{(13)}_h$ displays a distinct IR behavior from that of $\mathcal{P}^{(22)}_h$ due to the decoupling of the first-order tensor and scalar fluctuations in Eq.~\eqref{premise}. 
The universal IR scaling of secondary GWs discussed in Ref. \cite{Cai:2019cdl} applies to a bilinear source, which does not lead to such a decoupling in our paper.

As discussed in Ref.~\cite{Ota:2022hvh}, the suppression of the primordial tensor mode may result from the effective mass of the tensor fluctuation introduced by the one-loop correction in the effective action.
When integrating out the scalar perturbations, we may write the equation of motion of the tensor fluctuations as follows,
\begin{align}
	h''_{\boldsymbol{k}}(\tau)+2\mathcal{H}h'_{\boldsymbol{k}}(\tau)+(k^2+m^2_{\rm eff})  h_{\boldsymbol{k}}(\tau)=0,
\end{align}
and $m^2_{\rm eff}$ is found to be positive in our case. 
This mass term introduces a decaying solution even for $k\tau \ll 1$, as far as $\tau^2m_{\rm eff}^2\gtrsim 1$.
The evolution of superhorizon tensor fluctuations in the separate Universe perspective will also be discussed in Ref.~\cite{Ota:2022full}.
We also show the comparison of our new result $\big|\mathcal{P}_h^{(13)}\big|/3$ and the standard second-order SIGWs $\mathcal{P}_h^{(22)}$ in the right panel of Fig.~\ref{RD}. 
Here, the dimensionless time variable $y_{\rm D}$ is taken as $4\times10^3$, and the amplitudes of the scalar and the primordial tensor spectra are chosen as $A_\zeta=10^{-2}$ and $\mathcal{P}^{(11)}_h=10^{-10}$, respectively. $\mathcal{P}_h^{(13)}$ overwhelms $\mathcal{P}_h^{(22)}$ in the IR region, so the one-loop correction to the tensor spectrum on the superhorizon scale is dominated by $\mathcal{P}_h^{(13)}$.

\subsection{Early Matter-dominated era}\label{MDsection}

\begin{figure*}[htp]
    \centering    
    \includegraphics[width=\linewidth]{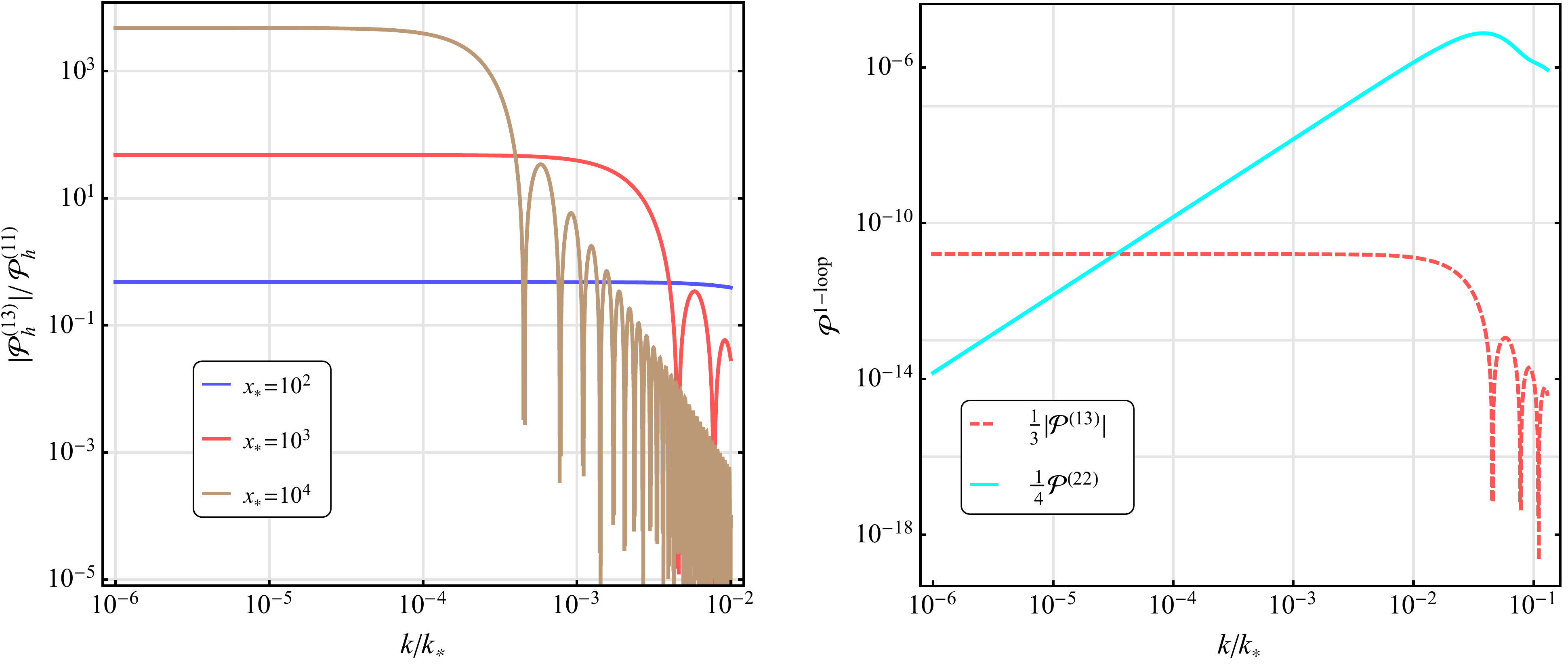}
    \caption{\label{MD} {\it Left}: The absolute value of the one-loop correction 
    $|\mathcal{P}^{(13)}_h|$ sourced  by the delta-function-like source during MD era. The vertical axis is normalized by the primordial tensor spectrum $\mathcal{P}^{(11)}_h$
    , and $A_\zeta$ is taken to be $10^{-4}$. Brown, red and blue solid lines denote different times $
    y\equiv k_*\tau=10^2, 10^3$ and $10^4$, respectively.
    The damping feature for $k/k_*\gtrsim y^{-1}$ implies the horizon entry.
    {\it Right}: 
    A comparison between two types of one-loop SIGWs power spectra. The cyan curve refers to $\frac{1}{4}\mathcal{P}_h^{(22)}$ while the red dashed line represents $\frac{1}{3}\big|\mathcal{P}_h^{(13)}\big|$. Here the dimensionless time $y$ is chosen to be $10^2$, $A_\zeta=10^{-4}$, and $\mathcal{P}_h=10^{-10}$.}
\end{figure*}
After inflation, the oscillation of massive fields effectively acts as pressureless dust.
Hence, there may be an early MD period.
The second-order SIGWs $\mathcal{P}_h^{(22)}$ during this era have been studied in Refs. \cite{Baumann:2007zm,Assadullahi:2009nf,Alabidi:2013lya,Kohri:2018awv,Gong:2019mui,Inomata:2019ivs,Inomata:2019zqy,Dalianis:2020gup}.
Since the gravitational potential is constant during the MD era, sizable SIGWs may be produced during the early MD era.
Therefore, we also investigate our new effect during the early MD era in this section.
The transfer functions during the MD era are given as
\begin{align}
    T_\phi(x)&=1 ,
    \\
    T_h(x)&=\frac{3j_1(x)}{x} .
\end{align}
Thus the linear gravitational potential is constant during the MD era, and the source function in Eq. \eqref{sourcef1} is greatly simplified as
\begin{align}
    &f_{h\phi\phi}(u,x)=T_\phi T_\phi\left(-\frac{40}{3}u^2T_h+48T_h-\frac{16}{3}u^2xT_h'\right) \nonumber
    \\           &=\frac{8\left[18+u^2(1-2x^2)\right]}{x^3}\sin{x}-\frac{8\left(u^2+18\right)}{x^2}\cos x,
    \label{finEMD}
\end{align}
and the iterative part is zero from Eq.~\eqref{eomphi2}.
Combining Eqs.~\eqref{finEMD} with~(\ref{P13}), we find the one-loop correction,
\begin{align}\label{p13md}
&\frac{\mathcal{P}^{(13)}_h(\tau,k)}{\mathcal{P}^{(11)}_h}=\int\frac{dp}{p}\frac{12\left(k\tau\cos{k\tau}-\sin{k\tau}\right)}{(k\tau)^6}
\notag \\
&\times \Big[k\tau\left(54-p^2\tau^2\right)\cos{k\tau}
\notag \\
&+\left(p^2\tau^2+18k^2\tau^2-54\right)\sin{k\tau} \Big] \times\left(\frac{3}{5}\right)^2\mathcal{P}_\zeta(p),
\end{align}
Let us consider a delta-function-like source in Eq. \eqref{deltasource}.
The final expression is similar to Eq.~(\ref{p13_final}), while the pre-factor $\mathcal{F}$ is given as
\begin{align} \label{F_matter}
    \mathcal{F}(k_*,k,\tau)\bigg|_{k\tau < 1} \simeq -\frac{2}{25}A_\zeta k_*^2\tau^2.
\end{align}
We show the power spectrum of the tensor modes in Fig.~\ref{MD} for $A_\zeta=10^{-4}$. 
The primordial power spectrum normalizes the final result. 
In the left panel, from top to bottom, brown, red, and solid blue lines denote different times $y=10^2, 10^3$ and $10^4$, respectively.
Since the scalar source is constant after horizon entry, it can continuously generate the tensor modes.
In the right panel of Fig.~\ref{MD}, we compare $\big|\mathcal{P}^{(13)}\big|$ and $\mathcal{P}^{(22)}$. 
Here we take $y=10^2$, the amplitude is $A_\zeta=10^{-4}$ and $\mathcal{P}^{(11)}_h=10^{-10}$.

Large enhancement of second-order SIGWs was also discussed in the previous literature, but there is a caveat about the induced tensor modes during the matter era.
Induced tensor modes contain not only the GWs, but also non-propagating modes during the matter era, and the latter is mainly amplified.
We cannot regard the non-propagating component as GWs since its energy contribution to the cosmic expansion is $a^{-2}$, i.e., the non-propagating part represents the curvature rather than the GWs. 
References~\cite{Inomata:2019zqy,Inomata:2019ivs} showed that the curvature is converted into the GWs if the transition from matter to radiation is faster than the oscillation time scale of scalar perturbations, and the curvature dilutes without sourcing the GWs if the transition is slow.
The same argument may apply to the present case, i.e., the final amplitude may be sensitive to the transition between two eras.
We leave further investigation for future work.

\subsection{Late Matter-dominated era}

The early matter dominance is hypothetical, but late-time matter dominance from recombination to dark-energy dominance is manifest.
We have already observed tiny and almost scale-invariant scalar fluctuations at that scale.
Does the observed curvature fluctuation change the CMB B-mode during the late-time matter era?
The subhorizon gravitational potential damps during the radiation era, so there is a natural UV cut-off in the loop integral.
For simplicity, let us assume  
\begin{align}\label{sisource}
    \mathcal{P}^c_\zeta=A_\zeta\Theta(k_{\rm eq}-k),
\end{align}
where $\Theta$ is the Heaviside step function, $A_\zeta\sim 2.1\times10^{-9}$, and $k_{\rm eq}\sim 0.01h{/\rm Mpc}$ is the horizon scale of the matter-radiation equality. 
In this case, using Eq.~\eqref{superposition}, the superhorizon spectrum is suppressed by a factor of 
\begin{align}
	\mathcal{F}(k,\tau) \sim -\frac{1}{25} A_\zeta\left(k_{\rm{eq}}\tau\right)^2.
\end{align} 
At recombination time $\tau_{\rm rec}\sim 300{~\rm Mpc}/h$, $k_{\rm eq}\tau_{\rm rec} =\mathcal O(1)$, so the correction is tiny, that is, $\mathcal P^{(13)}_h/\mathcal P^{(11)}_h=\mathcal O(A_\zeta)$.
Therefore, the CMB polarization from recombination will remain unchanged.
Reionization also introduces low-$\ell$ B modes with less lensing contamination.
The reionization time is given as $\tau_{\rm reio}\sim 4000{~\rm Mpc}/h$, so we find $\mathcal P^{(13)}_h/\mathcal P^{(11)}_h = \mathcal O(10^3A_\zeta)$, which could also be too small for the experiments.
Therefore, we conclude that we will not see a reduction in the CMB polarization in the present case.
However, we only consider the linear evolution of the scalar fluctuation in the late matter era.
The nonlinear evolution of cosmological perturbations plays an essential role during the matter era.
Hence including enhancement due to the nonlinearity in the above estimation will be interesting.

\section{Collaborative multi-frequency GW experiments for PBH detection}
\label{exp}

The direct measurement of stochastic GW background may tell us the amplitude of the small-scale curvature perturbations $A_\zeta$ and the mass distribution of PBHs in future observations. 
Then, we point out that low-frequency GWs may be secondarily reduced by large $A_\zeta$, which could be tested in the next-generation CMB experiments. For example, CMB-S4 and LiteBIRD, next-generation ground-based and space-based experiments, are expected to reach an upper limit of $r<0.001$
~\cite{CMB-S4:2020lpa, LiteBIRD:2020khw}.
Thus, combining the multi-frequency GW experiments range from $10^{-15}$ to $10^4$ Hz, from CMB polarization to the LIGO/Virgo, should be crucial to discuss PBH formation theories.

\begin{figure*}[ht]
	\centering
	\includegraphics[width=0.355\textheight]{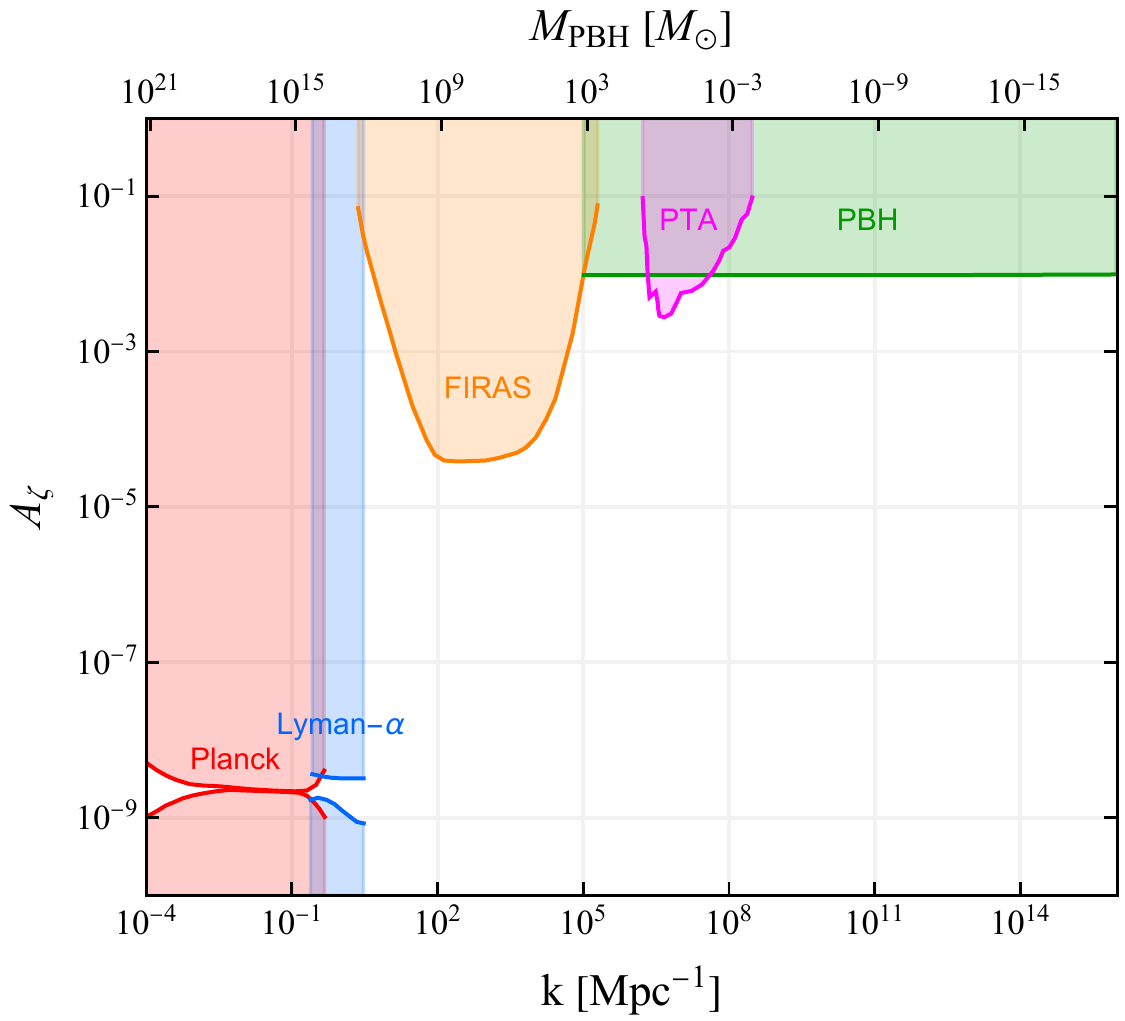}
	\includegraphics[width=0.355\textheight]{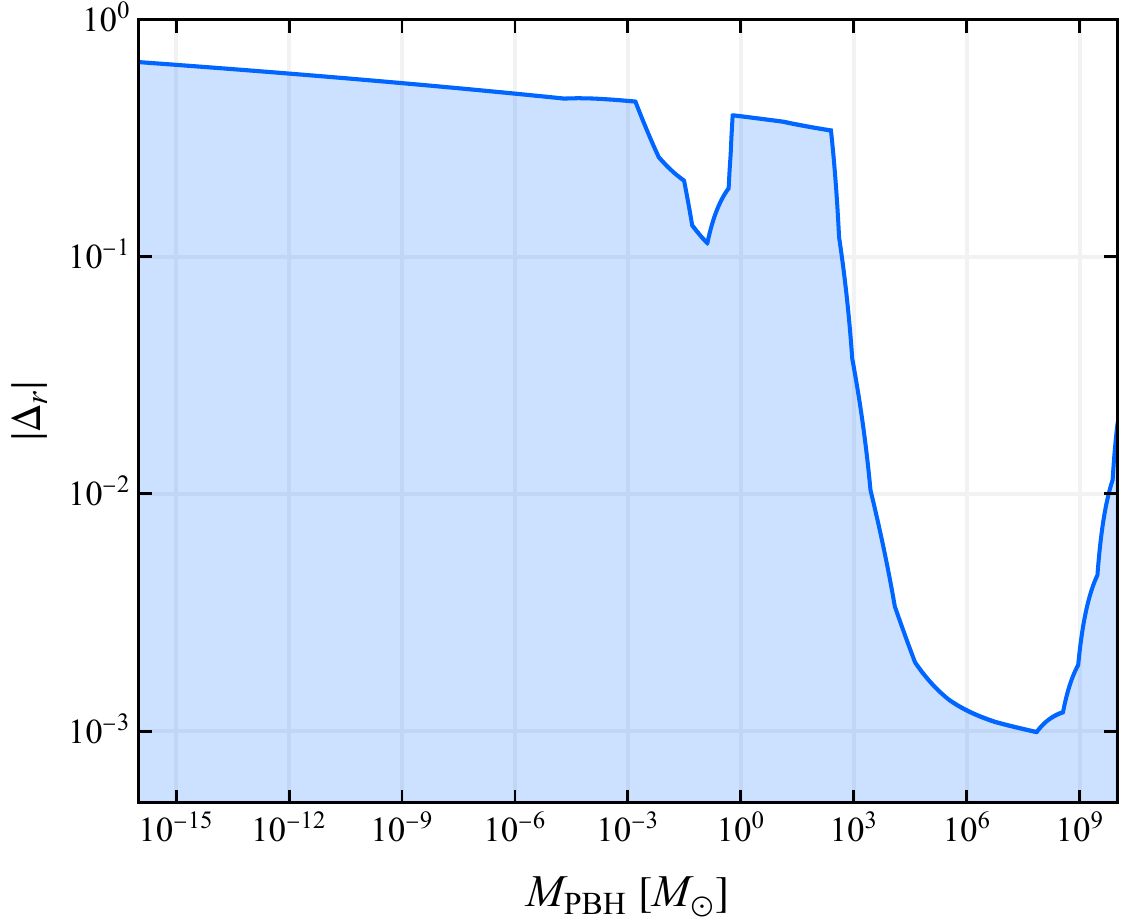}
	\caption{ {\it Left}: The current constraints on the curvature perturbations on different scales, including the Planck
	\cite{Planck:2018jri} (red), Lyman-$\alpha$ forest \cite{Bird:2010mp} (blue), FIRAS CMB spectral distortion \cite{Fixsen:1996nj} (orange) and PTA constraint on the standard SIGW $\mathcal{P}^\text{(22)}_h$ \cite{Byrnes:2018txb} (magenta). We also present constraint from PBH abundance account for
	dark matter~\cite{Inomata:2018epa, Green:2020jor} (green) with a conservative value $A_\zeta = 10^{-2}$. Similar plots can also be found in Refs.~\cite{Byrnes:2018txb, Inomata:2018epa, Green:2020jor}.
	The lower horizontal axis corresponds to the peak scale $k_*$ for the delta-function-like source. {\it Right}: The upper limit on tensor-to-scalar ratio variation $|\Delta_r|$ in terms of the monochromatic PBH mass $M_\text{PBH}$.
	The shadow refers to the parameter space allowed by the left panel.}	
	\label{fig:observation}
\end{figure*}

The accumulated effect of the one-loop correction $\mathcal{P}^{(13)}_h$ from the RD era ceases at $\tau_{\rm D}$ since scalar fluctuations below the diffusion scale $k_{\rm D}$ is exponentially suppressed as discussed in Sec.~\ref{RDsection}. The one-loop calculation during the MD era requires further investigation of nonlinear dynamics, so we limit our quantitative arguments to the RD era. In our case, the tensor-to-scalar ratio on the CMB scale is written as
\begin{equation}
r \equiv { 2 \Big( {1\over3} \mathcal{P}_h^\text{(13)}(\tau_{\rm D}) + \mathcal{P}^{(11)}_h \Big) \over A_\zeta^\text{CMB} } 
= r^\text{(11)} ( 1 + \Delta_r ),
\end{equation} 
where $\mathcal{P}^\text{(13)}_h$ can be calculated in Eq. \eqref{p13_final} and $A_\zeta^\text{CMB} \simeq 2.1 \times 10^{-9}$ \cite{Planck:2018jri}. The factor $2$ accounts for two polarizations of tensor modes.
Note that $r$ is evaluated at the pivot scale $k_s = 0.05 ~\text{Mpc}^{-1}$ for CMB observation. $r^\text{(11)} \equiv 2 \mathcal{P}^\text{(11)}_h/A_\zeta^\text{CMB}$ is the commonly-used definition of the tensor-to-scalar ratio for PGWs, and it is straightforward to see that
\begin{equation} \label{deltar}
\Delta_r = {1\over3} {\mathcal{P}^\text{(13)}_h(\tau_{\rm D}) \over \mathcal{P}^{(11)}_h}
\simeq  A_\zeta\left[ 4-4.8 \log (k_*\tau) \right] ,
\end{equation} 
which is calculated by Eq. \eqref{Famp}. From the above expression, $\Delta_r$ depends on $k_*$ and $A_\zeta$. 
At present, $A_\zeta$ is loosely constrained as shown in the left panel of Fig.~\ref{fig:observation}. 
However, we expect that PIXIE like spectral distortion experiments~\cite{Chluba:2019nxa} and GW experiments including SKA, LIGO and BBO \cite{Inomata:2018epa, Chluba:2019nxa}, will significantly improve the the upper bounds~(see Fig.~1 in Ref. \cite{Green:2020jor} or Fig.~4 in Ref. \cite{Inomata:2018epa} for details.).
Also, one can relate the peak scale $k_*$ to the formation masses of PBHs by using the horizon-mass approximation \cite{Sasaki:2018dmp}, 
\begin{align}
M_\text{PBH} &\simeq \gamma M_H
\notag \\ 
&\simeq M_\odot \left( \frac{\gamma}{0.2} \right) \left( \frac{g_{\text{form}}}{10.75} \right)^{-\frac16} \left( {k_* \over 1.9 \times 10^6~\text{Mpc}^{-1}} \right)^{-2},	
\end{align}
where $\gamma \simeq 0.2$, $g_{\text{form}} \simeq 106.75$ and $M_\odot$ is the solar mass. Hence, with the constraints on $A_\zeta$ shown in the left plot in Fig. \ref{fig:observation}, we can obtain upper limits on $|\Delta_r|$ for each PBH mass $M_\text{PBH}$.

The plot shows that the lighter PBHs reduce PGW more. 
Two concaves are due to the present constraints on $A_\zeta$ from FIRAS and PTA in the left panel in Fig.~\ref{fig:observation}. 
$|\Delta_r|$ in Fig.~\ref{fig:observation} may exceed the unity for a certain small $M_\text{PBH}$, implying the ignorance of the higher-order nonlinear terms or that we cannot trust perturbative analysis anymore because the auto tensor spectrum including all corrections must be non-negative.

A remaining issue is gauge dependence.
Tensor fluctuations at nonlinear order are generally gauge-dependent; therefore, comparing theory and observations is not straightforward.
Several works suggested that the induced GWs are physically well-defined only in the subhorizon scale, and the induced GWs are gauge independent in that limit~\cite{Inomata:2019yww,DeLuca:2019ufz,Domenech:2020xin,Ota:2021fdv}.
However, the loop effect we discussed is manifest at the superhorizon scale; that is, we consider the nonlinearity in the superhorizon tensor modes.
We cannot distinguish the third-order tensor fluctuations from the linear ones once the source disappears, and the evolution afterward is linear.
Therefore, the same solution does not apply to the gauge issue of the superhorizon corrections.
However, as we work in the same gauge condition for the rest of cosmic history, observational predictions such as the CMB polarization should be consistent.
We will further investigate the gauge dependence of the loop effect in future work.

\section{Conclusions}
\label{conclusion}

The nonlinear interaction of cosmological perturbations secondarily induces tensor fluctuations or GWs.
Such secondary GWs are attracting growing attention as we indirectly test the PBH formation theories via future GW measurements.
Recent works mostly considered the one-loop auto-power spectrum of second-order induced tensor modes.
This paper identified a missing one-loop contribution from the cross-power spectrum of first- and third-order tensor modes.
We computed the third-order tensor fluctuation sourced by a tensor and two scalar perturbations, including higher-order nonlinear interactions and iterative solutions.
Assuming a primordial tensor mode and enhanced delta-function-like scalar fluctuation in a typical PBH formation scenario, we found that the new one-loop correction is \textit{scale-invariant} and \textit{negative} in the superhorizon region.
Hence, short-scale large scalar fluctuations may significantly reduce the superhorizon primordial tensor power spectrum.
Suppose that the recent LIGO/Virgo events are explained by tens-solar-mass PBHs generated by a sharp peak of scalar fluctuations with $A_\zeta \sim 10^{-2}$ at $k_* \sim 10^{5}h/{\rm Mpc}$, we showed that the tensor power spectrum at the CMB scale reduces by at most 35\%.
Hence, the polarization B-mode might not be observed because the secondary effect of PBH formation reduced the original tensor spectrum. 
In a hypothetical early MD era, the reduction effect is more sensitive to the scalar amplitude since the gravitational potential is constant, implying that the loop expansion may easily fail.
Hence, a detailed loop analysis will be required for further quantitative predictions in MD eras.

\begin{figure}[htp]
	\centering
	\includegraphics[width=\linewidth]{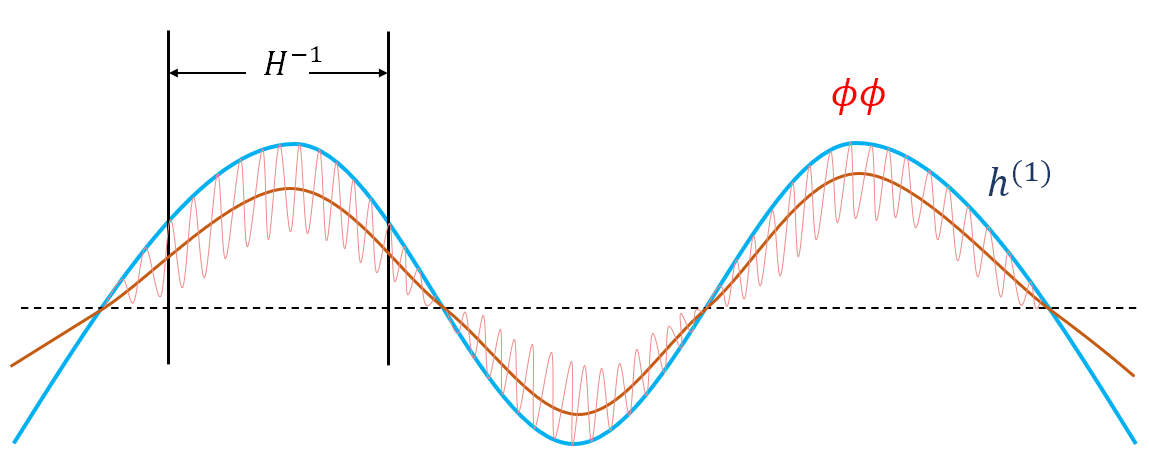}
	\caption{Modulation of the superhorizon tensor fluctuation is illustrated. The blue sine curve means the tensor fluctuations whose wavelength is larger than the Hubble scale $H^{-1}$. The thin orange curve means the amplitude modulation introduced by the subhorizon scalar fluctuations coupled to the first-order tensor fluctuations. When integrating out the subhorizon scalar fluctuations, we obtain the reduced tensor fluctuations denoted by the thick orange curve.}
	\label{fig:modulation}
\end{figure}

The new IR behavior greatly differs from the case of the second-order tensor auto-power spectrum since the causally generated second-order tensor fluctuations are never correlated over the superhorizon scale.
Then, does the scale-invariant reduction violate causality?
The new third-order correction is not a production of GWs from zero but a shift of the existing linear tensor modes via the Fourier mode coupling at nonlinear orders.
Equation~\eqref{premise} implies that the third-order correction is the amplitude modulation of superhorizon first-order tensor fluctuations by the subhorizon scalar fluctuations as illustrated in Fig.~\ref{fig:modulation}.
References~\cite{Pajer:2012vz,Ganc:2012ae,Ota:2016mqd,Ota:2014iva}
 discussed similar effects for CMB spectral distortion anisotropies in the presence of primordial non-Gaussianity.
They found that the local non-Gaussinaity introduces the Fourier mode coupling between super- and sub-horizon modes at third-order. Then the secondarily generated spectral distortions are correlated over the superhorizon scale without violating causality.
The superhorizon evolution of other cosmological perturbations due to the primordial non-Gaussianity is also discussed in Ref.~\cite{Ota:2017jte,Naruko:2015pva,Ota:2020vfn}. 
Our mechanism is essentially the same as this, while non-Gaussianity is naturally introduced by nonlinearity in the Einstein equation.
The variation of the superhorizon tensor mode will also be discussed with the separate Universe formalism in Ref.~\cite{Ota:2022full}.
Similar IR dependence was also found in the general relativistic correction of the matter power spectrum in Ref.~\cite{Jeong:2010ag} in the context of large-scale-structure, while the correction is tiny for $\zeta\sim10^{-5}$ in that work.

One of the authors recently claimed that the one-loop inflationary tensor power spectrum might be scale-invariantly enhanced or reduced due to a subhorizon resonant spectator scalar field~\cite{Ota:2022hvh}.
Their reduction effect is due to the Born approximation for the fourth-order interaction Hamiltonian, which corresponds to the third-order source in the equation of motion in our analysis.
The scale-invariant enhancement comes from the iterative correction of one-loop diagram (a1) in Fig.~\ref{fig:detail}. 
The iterative correction during inflation can be amplified and dominant when the scalar Green function is also enhanced for a nontrivial background.
We do not expect similar enhancement during radiation or matter eras.
Hence, the iterative correction could be comparable to the reduction effect in our case.
In addition, the inflationary loop correction is enhanced when amplifying the spectator fields without varying the slow-roll parameter since the interaction Hamiltonian during inflation is slow-roll suppressed.
Therefore, the inflationary one-loop correction is not necessarily enhanced for an arbitrary PBH formation scenario.  
However, the new significant reduction effect discussed in this paper always appears once $\zeta$ is amplified.
Hence, PBH formation and reduction of primordial GWs may be two sides of the same coin, suggesting that combining GW detectors at all scales is indispensable!

\section*{Acknowledgments}

CC, HYZ, and YHZ thank the Particle Cosmology Group at University of Science and Technology of China during their visits. AO would like to thank Keisuke Inomata, Misao Sasaki and Yi Wang for useful discussions. We use the Mathematica package MathGR \cite{Wang:2013mea} in this work. This work is supported in part by the National Key R\&D Program of China (No. 2021YFC2203100). The authors are supported by the Jockey Club Institute for Advanced Study at The Hong Kong University of Science and Technology. HYZ is supported in part by a grant from the RGC of the Hong Kong SAR, China (No.~16303220).

\appendix

\section{Iterative solutions in the IR region}
\label{app3}

In this appendix, we discuss the IR scaling of the three source terms we ignored in the main text.
The first contribution arises from 
\begin{align}
S^{(3)}_{\phi h_{\phi h},ij}=12\left[\phi h^{(2)}_{ij}{}''+\left(2\mathcal{H}\phi+\phi'\right)h^{(2)}_{ij}{}'\right].
\label{eq:S3pph}
\end{align}

The second-order tensor fluctuation in the above source is obtained by integrating
\begin{align}
h^{(2)}_{ij}{}''+2\mathcal{H}h^{(2)}_{ij}{}'-\nabla^2 h^{(2)}_{ij}=8\phi'h^{(1)}_{ij}{}'+8\phi\nabla^2 h^{(1)}_{ij}.
\label{eq:h2hphi}
\end{align} 
When substituting the solution of Eq.~\eqref{eq:h2hphi} into Eq.~\eqref{eq:S3pph}, we find the linear tensor fluctuation in the third-order source always appears with derivative operators.
Then, from Eq.~\eqref{premise}, those derivative operators turn into the external momentum when cross-correlating with the linear field.
Hence, these terms vanish in the IR region.

Secondly, we find the following source:
\begin{align}
    &S^{(3)}_{h\phi_{\phi\phi},ij}=
    2h^{(1)}_{ij}\left(\nabla^2\Psi^{(2)}-\nabla^2\Phi^{(2)}\right)
    \notag 
    \\
    &+6\left(\Phi^{(2)}+\Psi^{(2)}\right)\nabla^2h^{(1)}_{ij} +3\left(\Phi^{(2)}{}'
    +3\Psi^{(2)}{}'\right)h^{(1)}_{ij}{}'
    \notag \\
    &
   +3\left(\partial^k \Psi^{(2)}-\partial^k \Phi^{(2)}\right)\left(\partial_j h^{(1)}_{k  i}+\partial_i h^{(1)}_{k  j}-\partial_k  h^{(1)}_{ij}\right).
   \label{Shphp}
\end{align}
As discussed above, derivatives of tensor fluctuations will vanish in the IR region.
In addition, the second-order scalar fluctuations in the first line of Eq.~\eqref{Shphp} reduces to the zero modes so that $\nabla^2= 0$.
Therefore, we may safely ignore this source.
Similarly, cross-correlating $S^{(3)}_{hh_{\phi\phi},ij}$ with the linear tensor modes, $h_{\phi\phi}$ reduces to the zero mode, which should always be zero from statistical isotropy of cosmological perturbations.

\begin{widetext}

\section{Analytical Result in RD} \label{app2}
In this Appendix, we provide the analytical expression for the kernel function generated from the source $S_{h\phi\phi}^{(3)}$ during RD era, which reads
{\scriptsize
\begin{align}
&I_{h,h\phi\phi}(u,x)=\nonumber\\
   &\frac{3}{70u^6x^2}\Bigg\{2 \sqrt{3} \left(53 u^2-168\right) \sin{2x}\,\text{Si}\left(\frac{2 u x}{\sqrt{3}}\right) u^5+2 \sqrt{3} \left(53 u^4-168 u^2+630\right) u^3 \log
   \left|\frac{ \sqrt{3} u-3}{\sqrt{3} u+3}\right| \sin{^2x}\nonumber\\
   &-36 \left(35 u^6-70 u^4+84 u^2+36\right) \text{Ci}(2 x) \sin ^2x+2
   \left(-53 \sqrt{3} u^7+315 u^6+168 \sqrt{3} u^5-630 u^4-630 \sqrt{3} u^3+756 u^2+324\right) \text{Ci}\left(\frac{2}{3} \left|\sqrt{3} u-3\right| x\right) \sin
   ^2x\nonumber\\
   &+2 \left(53 \sqrt{3} u^7+315 u^6-168 \sqrt{3} u^5-630 u^4+630 \sqrt{3} u^3+756 u^2+324\right) \text{Ci}\left(\frac{2}{3}(\sqrt{3} u+3) x\right)
   \sin^2x\nonumber-18 \left(35 u^6-70 u^4+84 u^2+36\right) \log \left|\frac{ u^2-3 }{3}\right|\sin^2x\\
   &+\frac{6\sin{x}}{x^6} \bigg[3 x \cos x
   \bigg(-70 u^4 x^4+84 u^2 x^4+36 x^4+378 u^2 x^2-18 x^2-3 \sqrt{3} u x \left((9 u^2-4) x^2+180\right) \sin\frac{2 u x}{\sqrt{3}}+810
   \nonumber\\
   &+2
   \left(2 (13 u^4-24 u^2-9) x^4+9 (9 u^2+1) x^2-405\right) \cos \frac{2 u x}{\sqrt{3}}\bigg)+\sin x \bigg(106 u^6
   x^6+330 u^4 x^6-36 u^2 x^6-420 u^4 x^4+126 u^2 x^4+54 x^4-2079 u^2 x^2-81 x^2
   \nonumber\\
   &+\sqrt{3} u x \left(2 \left(26 u^4+57 u^2-18\right) x^4+9 \left(79 u^2+6\right)
   x^2+1620\right) \sin\frac{2 u x}{\sqrt{3}}-3 \left(2 \left(26 u^4+30 u^2+9\right) x^4-9 \left(17 u^2+3\right) x^2-810\right) \cos\frac{2
   u x}{\sqrt{3}}-2430\bigg)\bigg]
   \nonumber\\
   &+\sin{2x}\left(53 \sqrt{3} u^7-315 u^6-168 \sqrt{3} u^5+630 u^4+630 \sqrt{3} u^3-756 u^2-324\right)\text{Si}\left(\left(2-\frac{2 u}{\sqrt{3}}\right) x\right)+18 \sin{2x} \left(35 u^6-70 u^4+84 u^2+36\right)\text{Si}(2x)\nonumber\\
   &-\sin{2x}\left(53 \sqrt{3} u^7+315 u^6-168 \sqrt{3} u^5-630 u^4+630 \sqrt{3} u^3+756 u^2+324\right)  \text{Si}\left(\frac{2}{3} \left(\sqrt{3}
   u+3\right) x\right)\Bigg\}.
\end{align}
}

\end{widetext}

\bibliography{apssamp}

\end{document}